%% file: main.tex
\documentclass[sigconf]{acmart}

\usepackage{balance}
\def\BibTeX{{\rm B\kern-.05em{\sc i\kern-.025em b}\kern-.08emT\kern-.1667em\lower.7ex\hbox{E}\kern-.125emX}}

\usepackage{booktabs} 
\usepackage{algorithm}
\usepackage{algpseudocode}
\usepackage{multirow}
\usepackage{caption}
\usepackage{subcaption}
\usepackage{verbatim}
\usepackage{setspace}
\usepackage{geometry}
\usepackage{amsmath}
\usepackage{amsfonts}
\newcommand{\model}{{FacT}}

\DeclareMathOperator*{\argmin}{arg\,min}

\begin{document}
\copyrightyear{2019} 
\acmYear{2019} 
\setcopyright{acmcopyright}
\acmConference[SIGIR '19]{Proceedings of the 42nd International ACM SIGIR Conference on Research and Development in Information Retrieval}{July 21--25, 2019}{Paris, France}
\acmBooktitle{Proceedings of the 42nd International ACM SIGIR Conference on Research and Development in Information Retrieval (SIGIR '19), July 21--25, 2019, Paris, France}
\acmPrice{15.00}
\acmDOI{10.1145/3331184.3331244}
\acmISBN{978-1-4503-6172-9/19/07}

\settopmatter{printacmref=true}
\fancyhead{}

\title{The FacT: Taming Latent Factor Models for Explainability \\with Factorization Trees}

\author{Yiyi Tao}
\affiliation{
\department{Department of Physics}
\institution{Peking University}
\city{Beijing 100871}
  \country{China}
 }
\email{taoyiyi1111@pku.edu.cn}

\author{Yiling Jia, Nan Wang, Hongning Wang}
\affiliation{
\department{Department of Computer Science}
    \institution{University of Virginia} \city{Charlottesville}
  \state{VA 22904}
  \country{USA}
 }
\email{{yj9xs, nw6a, hw5x}@virginia.edu}

\fancyhead{}
\begin{abstract}

Latent factor models have achieved great success in personalized recommendations, but they are also notoriously difficult to explain. In this work, we integrate regression trees to guide the learning of latent factor models for recommendation, and use the learnt tree structure to explain the resulting latent factors. Specifically, we build regression trees on users and items respectively with user-generated reviews, and associate a latent profile to each node on the trees to represent users and items. With the growth of regression tree, the latent factors are gradually refined under the regularization imposed by the tree structure. As a result, we are able to track the creation of latent profiles by looking into the path of each factor on regression trees, which thus serves as an explanation for the resulting recommendations. Extensive experiments on two large collections of Amazon and Yelp reviews demonstrate the advantage of our model over several competitive baseline algorithms. Besides, our extensive user study also confirms the practical value of explainable recommendations generated by our model.
\end{abstract}


\maketitle

\input{intro}
\input{relwork}
\input{method}
\input{exp}
\input{userstudy}

\section{Conclusions and Future Works}
In this work, we seamlessly integrate latent factor learning with explanation rule learning for explainable recommendation. The fidelity of explanation is guaranteed by modeling the latent factors as a function of explanation rules; and the quality of recommendation is ensured by optimizing both latent factors and rules under a recommendation based metric.
Offline experiments and user studies have shown the effectiveness of our model in both aspects.

Our current work has much room for further improvement. Instead of using a set of single threshold predicates, we can introduce more complex forms, such as nonlinear function, for better explainability. Besides, \model{} is based on basic matrix factorization; but it is not limited to this form of latent factor models. We plan to develop other hybrid factorization models such as tensor factorization to integrate sentiment analysis with the rules. Last, our model only uses templates to generate explanations, we believe that using features as key words and retrieving sentences from items reviews will definitely generate more natural explanations.  

\section{Acknowledgement}
We want to thank Aobo Yang for his invaluable help in user study. Also, we thank the anonymous reviewers for their insightful comments. This paper is based upon work supported by the National Science Foundation under grant IIS-1553568 and CPS-1646501.

\input{main.bbl}


\end{document}

%% file: intro.tex
\section{Introduction}
Recommender systems have achieved great success in feeding the right content to the right user \cite{resnick1997recommender,breese1998empirical,sarwar2001item,Koren2009,Pazzani2007}. 
However, the opaque nature of most deployed recommendation algorithms, such as latent factor models \cite{Koren2009}, eagerly calls for transparency, i.e., explaining how/why the customized result is presented to a user \cite{Wang2018,Zhang2014,DBLP:journals,Ren:2017:SCV}. Previous research has shown that explanations help users make more accurate decisions \cite{bilgic2005explaining}, improve their acceptance of recommendations \cite{herlocker2000explaining}, and also increase their trust in the system \cite{sinha2002role}. Moreover, user studies find that users desire explanations of the personalized results - a survey of users of one popular movie recommendation system showed that over 86\% of those surveyed wanted an explanation feature \cite{herlocker2000explaining}.

We argue that the most important contribution of explanations in a recommender system is not to persuade users to adopt customized results (i.e., promotion), but to allow them to make more informed and accurate decisions about which results to utilize (i.e., satisfaction) \cite{bilgic2005explaining}. If users are persuaded to accept recommended results that are subsequently found to be inferior, their confidence and trust in the system will rapidly deteriorate \cite{herlocker2000explaining,sharma2013social}. Hence, the fidelity of explanations becomes a prerequisite for explainable recommendations to be useful in practice.

However, the fidelity of explanation and the quality of recommendation have long been considered as irreconcilable \cite{abdollahi2017using}: one has to trade recommendation quality for explanation. For example, it is believed that content-based collaborative filtering algorithms are easy to explain, as their underlying recommendation mechanism is straightforward. 
But due to their limited recommendation quality, the utility of such type of explanations is thus restricted. 

On the other hand, latent factor models \cite{Koren2009,rendle2010factorization} provide the most promising empirical performance in modern recommender systems, but they are hard to explain due to their complicated statistical structure. Various solutions have been proposed to \emph{approximate} the underlying recommendation mechanism of latent factor models for explanation. For example, Abdollahi and Nasraoui consider the most similar users and/or items in the learnt latent space as the explanation \cite{abdollahi2017using}. Phrase-level sentiment analysis is incorporated into latent factor learning for explanation, which maps users' feature-level opinions into the latent space and finds the most related features to the users and recommended items as explanations \cite{Wang2018,Zhang2014}. Similarly, topic models are introduced to model user-generated review content together with the latent factors for explainable recommendation \cite{McAuley2013,wang2011collaborative,Ren:2017:SCV}. However, to what extent these approximated  explanations comply with the learnt latent factor models is unknown, i.e., no guarantee in explanation fidelity. 

We believe the tension between recommendation quality and explanation fidelity is not necessarily inevitable; and our goal is to attain both by optimizing the recommendation in accordance with the designed explanation mechanism. In this work, we aim at explaining latent factor based recommendation algorithms with rule-based explanations. Our choice is based on the facts that 1) latent factor models have proved their effectiveness in numerous practical deployments \cite{Koren2009,rendle2010factorization}, and 2) prior studies show that rule-based explanations are easy to perceive and justify by the end-users \cite{clancey1983epistemology}. 
As the latent factors are not learned by rules, it is hard to craft any rules to explain the factors afterwards. Hence, we propose to integrate the rule-based decision making into the learning of latent factors. More specifically, we treat the latent factors as a function of the rules: based on different outcome of the rules, the associated groups of users and items should be routed to the designated latent factors, which are then optimized for recommendation. Due to similar characteristics shared by each group of users/items created by the learnt rules, the descriptive power of the learnt \emph{group-level} latent factors is enhanced, and the data sparsity problem in individual users/items could be substantially alleviated by this group-level latent factor learning. 

\begin{figure}
\vspace{-2mm}
\includegraphics[width=0.9\linewidth]{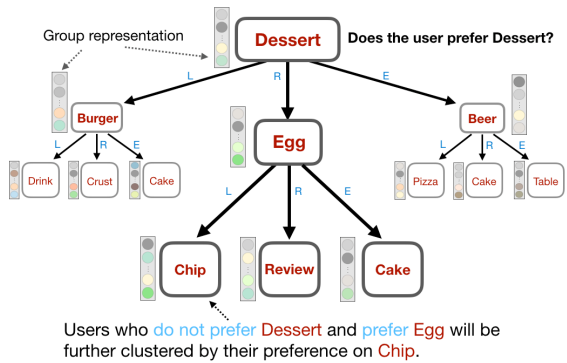}
\caption{An example user tree: Top three levels of our FacT model learnt for restaurant recommendations.}
\label{fig-tree}
\vspace{-4mm}
\end{figure}

More specifically, we format the explanation rules based on feature-level opinions extracted from user-generated review content, e.g., whether a user holds positive opinion towards a specific feature. The rules are extracted by inductive learning on the user side and item side separately, which form a user tree and an item tree. We alternate the optimization between  tree construction and latent factor estimation under a shared recommendation quality metric. An example of user tree is shown in Figure \ref{fig-tree}. For instance, according to the figure, if two users both expressed their preference of ``\textit{burger}'' in their reviews, they should be assigned to the same node on the user tree to share the latent user factors; accordingly, if two restaurants receive similar negative comments about their ``\textit{cleanliness}'', they should appear in the same node on the item tree. 
In testing time, the learnt user and item factors are used for recommendation as in standard latent factor models, and the rules that lead to the chosen user and item factors are output as explanations: e.g., ``\textit{We recommend item X because it matches your preference on burger and cleanness of a restaurant.}''  

Extensive experiment evaluations on two large sets of reviews, i.e., Amazon reviews for product recommendation and Yelp reviews for restaurant recommendation, demonstrate improved quality in recommendation and  explanation from our algorithm, compared with a set of state-of-the-art explainable recommendation algorithms. In particular, we perform serious user studies to investigate the utility of our explainable recommendation in practice, in both warm-start and cold-start settings. Positive user feedback further validates the value of our proposed solution. 

%% file: relwork.tex
\section{Related Works}
Various studies show that accurate explanation improves transparency of automated recommender systems \cite{sinha2002role,tintarev2011designing}, helps users make more informed decisions \cite{herlocker2000explaining,tintarev2007survey}, and thus increases recommendation effectiveness, user satisfaction and trust \cite{bilgic2005explaining,Tintarev:2007:ER}. 
There has been a substantial body of research on explainable recommendation. Broadly speaking, we categorize the existing explanation methods into neighbor-based and feature-based categories. Both of them however suffer from the trade-off between recommendation quality and explanation fidelity to different extents. 

The neighbor-based explanation methods root in content-based collaborative filtering \cite{sarwar2001item,breese1998empirical}. As the recommendations are made directly by measuring similarities between users and/or items, explaining the recommended results becomes straightforward. 
For example, Herlocker et al. proposed 21 types of explanation interfaces for a collaborative filtering system \cite{herlocker2004evaluating} and found a histogram showing the ratings from similar users was the most persuasive. Sharma and Cosley \cite{sharma2013social} conducted user studies to investigate the effect of social explanations, e.g., ``X, Y and 2 other friends like this.'' But the unsatisfactory recommendation quality limits the utility of provided explanations at the first place. This type of explanation has also been used in latent factor based collaborative filtering algorithms, where the similarity is measured in the learnt latent space \cite{abdollahi2017using}. However, as the latent space is not constructed for explanation, there is no guarantee that such type of explanations will comply with the recommendations. 

Feature-based explanation methods introduce information beyond classical dyadic interaction between users and items, such as user ratings and clicks. Earlier work in this category uses meta-data of items for explanation. For instance, Tintarev et al. \cite{Tintarev:2007:ER} use genre, director and cast to explain movie recommendations. Bilgic and Mooney \cite{bilgic2005explaining} extracted keywords from recommended books as explanations. But such explanations are heavily item-focused, and therefore independent of the recommendation algorithms. Their fidelity is often questionable. 
Later works in this category integrate feature representation learning with recommendation model learning, with the hope that the provided explanations can best correlate with the recommendations. For example, in \cite{Zhang2014,Wang2018}, phrase-level sentiment analysis is first used to extract users' feature-level descriptions of the items, and joint matrix or tensor factorization is then performed to map users, items and features onto the same latent space. The explanations are created by looking for the most related features to the user and recommended items in the learnt latent space, which is essentially neighbor-based explanation. But as the feature representation is learnt jointly with user and item representations, this type of explanations is believed to be more relevant and informative. 
Recently, neural attentive models are also developed to directly rank user reviews for explanation \cite{chen2018neural}. However, as the feature representation learning is only introduced as a companion task of recommendation learning, there is still no guarantee on the fidelity of provided explanations. 

The idea of providing rule-based explanations was popularized in the development of expert systems \cite{neches1985enhanced,wick1992reconstructive}. For example, MYCIN \cite{van1984emycin}, a rule-based reasoning system, provides explanations by translating traces of rules followed from LISP to English. A user could ask both why a conclusion was arrived at and how much was known about a certain concept. But since modern recommender systems seldom use rule-based reasoning, there is very little research on explaining latent factor models with rules. We propose to embed latent factor learning under explanation rule learning, by treating the latent factors as a function of rules, such that the generated explanations can strictly adhere to the provided recommendations. On a related note, a existing work \cite{WangXiang2018TEM} uses gradient boosting decision trees (GBDT) to learn rules from the reviews and incorporate rules into an attention network. But it only uses the rules as the input of embedding models and thus isolates the learning of tree structure and embeddings. Some systems \cite{Zhou:2011,Sun:2013} combine decision tree learning with matrix factorization to extract a list of interview questions for solving the cold-start problem in recommendation. But the rules are only built on the user side with their rating responses to items, i.e., the same as matrix factorization's input; it thus cannot provide any explanation to the recommended items.  

%% file: method.tex
\section{Methodology}
We elaborate our solution for joint latent factor learning and explanation rule construction in this section. Briefly, we model the latent factors for both users and items as a function of the rules: users who provide the same responses to the same set of rules would share the same latent factors, and so do the items. The predicates of rules are selected among the text features extracted from user-generated reviews. For example, whether a specific user expressed his/her preference on a particular feature in reviews. And the rules are constructed by recursive inductive learning based on previously selected predicate's partition of users and items. To reflect the heterogeneity between users and items, we construct rules for users and items separately. As a result of rule induction, the latent factors for users and items are organized in a decision tree like structure accordingly, where each node on the tree represents the latent factors for the group of users or items routed to that node. We alternate the optimization of the explanation rule construction and latent factor learning under a recommendation quality based metric. Hence, we name our solution as \emph{Factorization Tree}, or \emph{\model{}} in short. 

We start our discussion with factorization based latent factor learning, which is the basic building block of \model{}. Then we provide details in rule induction based on the learnt latent factors. Finally, we integrate these two learning components with an alternative optimization procedure. 

\subsection{Latent Factor Learning}
Latent factor models \cite{Koren2009,rendle2010factorization} have been widely deployed in modern recommender systems. The idea behind this family of solutions is to find vectorized representations of users and items in a lower dimensional space, which capture the affinity between users and items. Various latent factor models have been developed, such as matrix/tensor factorization \cite{Koren2009} and factorization machines \cite{rendle2010factorization}. Our \model{} is independent of the choice of latent factor models, as it treats the latent factor learning as a sub-routine. We apply matrix factorization in this paper due to its simplicity. Later in Section \ref{sec_rule_learning}, we will discuss how \model{} can be flexibly extended to other latent factor models. 

Formally, denote $\mathcal{U} = \{u_1,u_2,...,u_m\}$ as a set of $m$ users, $\mathcal{V} = \{ v_1, v_2,..., v_n \}$ as a set of $n$ items, and $r_{ij}$ as an observed rating from user $i$ to item $j$. 
The goal of latent factor learning is to associate each user and each item an $d$ dimensional vector respectively, i.e., $u_i\in \mathbb{R}^d$ and $v_j\in \mathbb{R}^d$, such that the inner product between user $i$'s factor and item $j$'s factor predicts the rating $r_{ij}$. The latent factors for all the users and items, denote as $U \in \mathbb{R}^{m \times d}$ and $V \in \mathbb{R}^{n \times d}$, can thus be learnt by minimizing their prediction error over a set of observed ratings $O = \{(i, j) |  r_{ij} \,is\,observed\}$ as follows,
\begin{equation}\label{point_obj}
  \mathcal{L}(U,V,O) = \min \limits_{U,V}\sum_{(i,j) \in O}(r_{ij} - u_i^\top v_j)^2.
\end{equation}
  
It is well accepted that recommendation is essentially a ranking problem \cite{Rendle:2009,yang2011collaborative}; however, the objective function introduced in Eq \eqref{point_obj} cannot fully characterize the need of ranking, i.e., differentiate the relative quality among candidates. To supplement information about relative item ranking into latent factor learning, Bayesian Personalized Ranking (BPR) loss \cite{Rendle:2009} has been popularly adopted to enforce pairwise ranking order. 
To realize the BPR loss, one needs to first construct a pairwise ordered set of items $D_i^o$ for each user $i$: $D_i^o := \{(j,l)| \,r_{ij} > r_{il}\}$, where $r_{ij} > r_{il}$ means that given the observations in $O$, either user i gives a higher rating to item $j$ than item $l$, or item $j$ is observed in user $i$'s rating history, while item $l$ is not. Then, the BPR loss can be measured on each user $i$ as:
\begin{equation*}
  \mathcal{B}(u_i, V, D_i^o) = \sum_{(j,l)\in D_i^o}\log\sigma(u_i^\top v_j- u_i^\top v_l)
\end{equation*}
where $\sigma(\cdot)$ is a logistic function. 

Putting together the pointwise rating prediction loss with the pairwise ranking loss, the latent factors for users and items can be learnt by solving the following optimization problem:
\begin{equation}
\label{eq_final_obj}
(\hat U, \hat V) = \argmin_{U, V} \mathcal{L}(U,V,O) - \lambda_b \sum_i \mathcal{B}(u_i, V, D_i^o)  + \lambda_u \|U\|_2 + \lambda_v \|V\|_2
\end{equation}
where $\lambda_b$ is a trade-off parameter to balance these two types of loss, $\|U\|_2$ and $\|V\|_2$ are L2 regularizations to control model complexity, and $\lambda_u$ and $\lambda_v$ are the corresponding coefficients. 
Eq \eqref{eq_final_obj} can be efficiently addressed by gradient-based optimization \cite{mairal2010online}. 
Once the user factors $U$ and item factors $V$ have been learnt, the recommendations for user $i$ can be generated by returning the top ranked items based on the predicted ratings $\hat r_{ij}=\hat u^\top_i \hat v_j$.

The premise behind the aforementioned learning procedure is that there is only a small number of factors influencing users' preferences, and that a user's preference vector is determined by how each factor applies to that user and associated items. But the factors are retrieved by solving a complex optimization problem (e.g., Eq \eqref{eq_final_obj}), which makes the resulting recommendations hard to explain. In \model{}, we embed latent factor learning under explanation rule construction, so that why a user or an item is associated to a particular latent factor can be answered by the matched rules, so do the generated recommendations.

\subsection{Explanation Rule Induction}
\label{sec_rule_learning}

In \model{}, we consider the latent factors as a function of explanation rules: the latent user factor $u_i$ for user $i$ is tied to the outcomes of a set of predicates applied to him/her, so does the latent item factor $v_j$ for item $j$. Based on different outcomes of the rules, the associated groups of users and items should be routed to the designated latent factors. At testing time, the activated predicates on user $i$ and item $j$ naturally become the explanation of this recommendation.

We select the predicates among the item features extracted from user-generated reviews. User reviews provide a fine-grained understanding of a user's evaluation of an item \cite{wang2010latent}. Feature-level sentiment analysis techniques \cite{Lu:2011:ACC} can be readily applied to reviews to construct a domain-specific sentiment lexicon. Each lexicon entry takes the form of (feature, opinion, sentiment polarity), abbreviated as $(f,o,s)$, and represents the sentiment polarity $s$ inferred from an opinionated text phrase $o$ describing feature $f$. Specifically, we label the sentiment polarity $s$ as +1 or -1, to represent positive or negative opinions. As how to construct a sentiment lexicon with phrase-level sentiment analysis is not the focus of this work, we refer interested readers to \cite{Lu:2011:ACC, Zhang2014} for more details.

The extracted item features become candidate variables for predicate selection. 
To compose predicates for explanation rule construction, we first need to define the evaluation of a single variable predicate on users/items according to their association with the item features. To respect the heterogeneity between users and items, we construct the predicates for users and items separately; but the construction procedures are very similar and highly related on these two sides.  

Denote $\mathcal{F} = \{ f_1, f_2,...,f_k \} $ as a set of $k$ extracted item features. 
Suppose feature $f_l$ is mentioned by user $i$ for $p^u_{il}$ times with a positive sentiment polarity  in his/her reviews and $n^u_{il}$ times with a negative sentiment polarity. We can construct a feature-level profile $F^u_i$ for user $i$, where each element of $F^u_i$ is defined as,
\begin{equation}
\label{eq_user_profile}
F^u_{il}= 
\left\{\begin{array}{rcl}
\emptyset,       &      & {\text{if}\quad p^u_{il} = n^u_{il} = 0,}\\
p^u_{il}+n^u_{il},       &      & \text{otherwise.}
\end{array} \right.
\end{equation}
Intuitively, $F^u_{il}$ is the frequency of user $i$ mentioning feature $f_l$ in his/her reviews, such that it captures the relative emphasis that he/she has given to this feature. And similarly, on the item side, denote $p^v_{jl}$ as the number of times that feature $f_l$ is mentioned in all user-generated reviews about item $j$ with a positive sentiment polarity, and $n^v_{jl}$ as that with a negative sentiment polarity, we define the feature-level profile $F^v_j$ for item $j$ as,
\begin{equation}
F^v_{jl}= 
\left\{\begin{array}{rcl}
\emptyset,       &      & {\text{if}\quad p^v_{jl} = n^v_{jl} = 0},\\
p^v_{jl}-n^v_{jl},       &      & \text{otherwise.}
\end{array} \right.
\end{equation}
Accordingly, $F^v_{jl}$ reflects the aggregated user sentiment evaluation about feature $f_l$ of item $j$.

Based on the feature-level user and item profiles, the evaluation of a single variable predicate can be easily performed by comparing the designated feature dimension in the user or item profile against a predefined threshold. For example, on the user side, if a predicate is instantiated with feature $f_l$ and threshold $t^u_l$, all users can have three disjoint responses to this predicate based on their $F^u_{il}$ values, i.e., $F^u_{il}\ge t^u_l$, or $F^u_{il}< t^u_l$, or $F^u_{il}$ is unknown. This gives us the opportunity to model the latent factors as a function of the explanation rules: based on the evaluation results of a predicate, we allocate the input users into three separate user groups and assign one latent vector per group. We should note that other forms of predicates are also applicable for our purpose, e.g., select a list of thresholds or a nonlinear function for one variable. For simplicity, we adhere to the form of single threshold predicates in this paper, and leave the more complex forms of predicates for future exploration.

Two questions remain to be answered: First, how to select the threshold for user-side and item-side predicate creation; and second, how to assign latent vectors for each resulting user/item group. We answer the first question in this section by inductive rule learning, and leave the second to the next section, where we present an alternative optimization procedure for joint rule learning and latent factor learning. In the following discussions, we will use user-side predicate construction as an example to illustrate our rule induction method; and the same procedure directly applies to item-side predicate construction.

Intuitively, an optimal predicate should create a partition of input users where the latent factors assigned to each resulting user group lead to  minimal recommendation loss defined in Eq \eqref{eq_final_obj}. This can be achieved by exhaustively searching through the combination of all item features in $\mathcal{F}$ and all possible corresponding thresholds. This seems infeasible at a first glance, as the combinatorial search space is expected to be large. But in practice, due the sparsity of nature language (e.g., Zipf's law \cite{newman2005power}), the mentioning of item features and its frequency in user reviews are highly concentrated at both user-level and item-level \cite{Wang2018}. Besides, feature discretization techniques \cite{dougherty1995supervised} can also be used to further reduce the search space.     

To perform the search for optimal predicate in an input set of users $\mathcal{U}_a$, we first denote the resulting partitions of $\mathcal{U}_a$ by feature $f_l$ and threshold $t^u_l$ as,
\begin{align}
\label{eq_branches}
L(f_l,t^u_l | \mathcal{U}_a) &= \{i|F^u_{il} \ge t^u_l, i \in \mathcal{U}_a\}, \nonumber\\
R(f_l,t^u_l | \mathcal{U}_a) &= \{i|F^u_{il} < t^u_l, i \in \mathcal{U}_a\}, \\
E(f_l,t^u_l | \mathcal{U}_a) &= \{i|F^u_{il} = \emptyset, i \in \mathcal{U}_a\}, \nonumber
\end{align}
and the set of possible threshold $t^u_l$ for feature $f_l$ as $T^u_l$. 
The optimal predicate on $\mathcal{U}_a$ can then be obtained by solving the following optimization problem with respect to a given set of item factors $V$,
\begin{align}\label{eq_rule_selection}
\small
 (\bar f_l, \bar t^u_l) =
 \!\!\argmin_{f_l\in\mathcal{F},t^u_l\in T^u_l} \!\min_{u_L, u_R, u_E} 
 &\mathcal{L}(u_L, V, O_L)\! - \!\lambda_b\!\! \sum_{i\in E(f_l,t^u_l)} \mathcal{B}(u_L, V, D^o_i) \nonumber \\ 
 +&\mathcal{L}(u_R, V, O_R)\! - \!\lambda_b\!\! \sum_{i\in R(f_l,t^u_l)} \mathcal{B}(u_R, V, D^o_i) \nonumber \\ 
 +&\mathcal{L}(u_E, V, O_E)\! - \!\lambda_b\!\! \sum_{i\in E(f_l,t^u_l)} \mathcal{B}(u_E, V, D^o_i) \nonumber \\
 +&\lambda_u (\|u_L\|_2+\|u_R\|_2+\|u_E\|_2)  
\end{align}
\normalsize
where $O_L$, $O_R$ and $O_E$ are the observed ratings in the resulting three partitions of $\mathcal{U}_a$, and $u_L$, $u_R$ and $u_E$ are the correspondingly assigned latent factors for the users in each of the three partitions. As users in the same partition are forced to share the same latent factors, the choice of text feature $f_l$ and corresponding threshold $t^u_l$ directly affect recommendation quality. In practice, considering each user and item might associate with different number of reviews, the size of user profile $F^u_i$ and item profile $F^v_j$ might vary significantly. Proper normalization of $F^u_i$ and $F^v_j$ can be performed, e.g., normalize by the total observation of feature mentioning in each user and item respectively. In this work, we follow \cite{dougherty1995supervised} for feature value normalization and discretization. 

Inside the optimization of Eq \eqref{eq_rule_selection}, a sub-routine of latent factor learning is performed to minimize recommendation loss induced by matrix factorization (as defined in Eq \eqref{eq_final_obj}) on the resulting partition of users. As we mentioned before, the choice of latent factor models does not affect the procedure of our predicate construction for \model{}, and many other recommendation loss metrics or latent factor models can be directly plugged into Eq \eqref{eq_final_obj} for explainablity enhancement. We leave this exploration as our future work. 

Our predicate construction can be recursively applied on the resulting user partitions $L(\bar f_l, \bar t^u_l | \mathcal{U}_a)$, $R(\bar f_l, \bar t^u_l | \mathcal{U}_a)$ and $E(\bar f_l, \bar t^u_l | \mathcal{U}_a)$ on the input user set $\mathcal{U}_a$ to extend a single variable predicate to a multi-variable one, i.e., inductive rule learning. The procedure will be terminated when, 1) the input user set cannot be further separated, e.g., all users there share the same user profile; or 2) the maximum depth has been reached. Starting the procedure from the complete set of users $\mathcal{U}$, the resulting set of multi-variable predicates form a decision tree like structure, which we refer as user tree in \model{} (as shown in Figure \ref{fig-tree}). On the user tree, each node hosts a latent factor assigned to all its associated users, and its path to the root node presents the learnt predicates for this node. The same procedure can be applied on the item side with a given set of user factors $U$ to construct item-specific predicates, i.e., item tree. 

Once the user tree and item have been constructed, explaining the recommendations generated by the latent factors becomes straightforward. Assume we recommend item $j$ to user $i$: we first locate user $i$ and item $j$ at the leaf nodes of user tree and item tree accordingly, extract their paths back to each tree's root node, and find the shared features on the paths to create feature-level explanations. As each branch on the selected path corresponds to a specific outcome of predicate evaluation, e.g., Eq \eqref{eq_branches}, we can add predefined modifiers in front of the selected features to further elaborate the associated latent factors. For example,  
\begin{itemize}
\item \textit{We recommend this item to you because its [good/excellent] [feature 1] matches with your [emphasize/taste] on [feature 1], and ...}
\item \textit{We guess you would like this item because of your [preference/choice] on [feature 1], and ...}
\end{itemize}
It is also possible that the number of shared features on the two paths is low, especially when the maximum tree depth is small. In this situation, one can consider to use the union of features on these two paths, and give higher priority to the shared features and those at the lower level of the trees, as they are more specific. Another possible way of explanation generation is to use the selected features to retrieve sentences from the corresponding item reviews \cite{chen2018neural}. But this approach is beyond the scope of this paper, and we leave it as our future work.  

\subsection{Alternative Optimization}
\label{sec:alternative}
The aforementioned procedure for explanation rule induction is intrinsically recursive and requires the availability of user factors for item tree construction and item factors for user tree construction. In this section, we will unify the learning of latent factors with tree construction to complete our discussion of \model{}.

Define the maximum rule length, i.e., tree depth, as $h$. We alternate rule induction by recursively optimizing Eq \eqref{eq_rule_selection} between user side and item side. At iteration $t$, we start induction from the complete user set $\mathcal{U}$ with the latest item factors $V_{t-1}$. For each pair of feature and threshold in the hypothesis space of Eq \eqref{eq_rule_selection}, we use gradient based optimization for learning latent factors according to Eq \eqref{eq_final_obj}. Once the induction finishes, we collect the latent user factors $U_t$ from the leaf nodes of the resulting user tree, and use them to execute the rule induction on the item side from the complete item set $\mathcal{V}$ to estimate $V_{t}$. This procedure is repeated until the relative change defined in Eq \eqref{eq_final_obj} between two consecutive iterations is smaller than a threshold, or the maximum number of iterations is reached. To break the inter-dependency between item tree and user tree construction, we first perform plain matrix factorization defined in Eq \eqref{eq_final_obj} to obtain the initial item factors $V_0$. We should note that one can also start with item tree construction from initial user factors $U_0$, but this does not change the nature and convergence of this alternative optimization. 

The above alternative optimization is by nature greedy, and its computational complexity is potentially high. When examine the optimization steps in Eq \eqref{eq_rule_selection}, we can easily recognize that the exhaustive search of item features and thresholds can be performed in parallel in each input set of users and items. This greatly improves the efficiency of rule induction. Besides, beam search \cite{ow1988filtered} can be applied in each step of predicate selection to improve the quality of learnt rules and factors, but with a cost of increased computation.   

One can realize that during the alternative optimization, only the latent factors learnt for the leaf nodes are kept for next round of tree construction and finally the recommendation, while the factors associated with the intermediate nodes are discarded. As the procedure of inductive rule learning can be considered as a process of divisive clustering of users and items, the intermediate nodes actually capture important information about homogeneity within the identified user clusters and item clusters. To exploit such information, we introduce the learnt latent factors from parent node to child node as follows,
\begin{equation*}
u_{L, z} = \tilde u_{L,z} + u_{z}, ~~
u_{R, z} = \tilde u_{R,z} + u_{z}, ~~\text{and}~~
u_{E, z} = \tilde u_{E,z} + u_{z},
\end{equation*}
where $u_{L, z}$, $u_{R, z}$ and $u_{E, z}$ are the  latent factors to be plugged into Eq \eqref{eq_rule_selection} for the three child nodes under parent node $z$, and $u_{z}$ is the factor already learnt for the parent node $z$. Intuitively, $\tilde u_{L,z}$, $\tilde u_{L,z}$ and $\tilde u_{L,z}$ can be considered as residual corrections added to the shared representation from parent nodes. Hence, the rule induction process becomes a recursive procedure of latent factor refinement. Without loss of generality, this recursive refinement can be applied to individual users and items on the leaf nodes of both user tree and item tree as well. If we refer the latent factors on the leaf node for individual users and items as personalized representations of users and items, those on the intermediate nodes could be considered as grouplized representations for the partition of users and items.  

%% file: exp.tex
\section{Experiments}
\label{sec_exp}
We performed a set of controlled experiments on two widely used benchmark datasets collected from Amazon\footnote{http://jmcauley.ucsd.edu/data/amazon/} and Yelp Dataset Challenge\footnote{https://www.yelp.com/dataset} to quantitatively evaluate our FacT model. We extensively compared \model{} against several state-of-the-art recommendation algorithms in both recommendation and explanation quality.

\subsection{Experiment Setup}
\noindent\textbf{$\bullet$ Preprocessing.} We utilize the restaurant businesses dataset from Yelp, and cellphones and accessories category dataset from Amazon in our evaluation. As discussed in \cite{Wang2018}, the dataset is quite sparse, e.g., 73\% users and 47\% items only have one review in Amazon dataset. To refine the raw data, we performed recursive filtering to alleviate the sparsity issue by taking two main steps: First, we preserve the features with frequency higher than a threshold; Second, we filter out the reviews that mention such features below another threshold. In the meantime, items and users that are associated with too few reviews were also removed. By fine tuning these different thresholds, we finally obtained two datasets with a decent number of users and items, whose statistics are shown in Table~\ref{tab:datasets}.

\begin{table}
  \caption{Basic statistics of evaluation datasets.}
  \label{tab:datasets}
  \vspace{-3mm}
  \begin{tabular}{|c|c|c|c|c|c|}
\hline
Dataset & \#users & \#items & \#features & \#opinions & \#reviews \\ \hline
Amazon  & 6,285  & 12,626 & 101       & 591      & 55,388   \\
Yelp    & 10,719 & 10,410 & 104      & 1,019     & 285,346  \\ \hline
\end{tabular}
\vspace{-3mm}
\end{table}

\begin{table*}[t]
\caption{Comparison of recommendation performance.}
\label{tab:baselines}
\vspace{-3mm}
\begin{tabular}{|c|l|l|l|l|l|l|l|l|c|}
\hline
\multirow{2}{*}{\begin{tabular}[c]{@{}c@{}}NDCG\\ @K\end{tabular}} & \multicolumn{8}{c|}{Amazon}                                            & \multirow{2}{*}{\begin{tabular}[c]{@{}c@{}}Improvement\\ best v.s. second best\end{tabular}} \\ \cline{2-9}
    & FMF    & MP     & NMF    & BPRMF   & JMARS  & EFM    & MTER   & FacT   &                                                                                             \\ \hline
10                                                                 & 0.1009 & 0.0961 & 0.0649 & 0.1185  & 0.1064 & 0.1109 & 0.1351 & {\bf 0.1482} & 9.70\%*                                                                                      \\ \hline
20                                                                 & 0.1331 & 0.1310 & 0.0877 & 0.1490  & 0.1348 & 0.1464 & 0.1653 & {\bf 0.1795} & 8.59\%*                                                                                      \\ \hline
50                                                                 & 0.1976 & 0.1886 & 0.1601 & 0.2070  & 0.1992 & 0.2056 & 0.2234 & {\bf 0.2367} & 5.95\%*                                                                                      \\ \hline
100                                                                & 0.2529 & 0.2481 & 0.2144 & 0.2669  & 0.2575 & 0.2772 & 0.2803 & {\bf 0.2869} & 2.35\%*                                                                                      \\ \hline
\multirow{2}{*}{\begin{tabular}[c]{@{}c@{}}NDCG\\ @K\end{tabular}} & \multicolumn{8}{c|}{Yelp}                                              & \multirow{2}{*}{\begin{tabular}[c]{@{}c@{}}Improvement\\ best v.s. second best\end{tabular}} \\ \cline{2-9}
& FMF    & MP     & NMF    & BPRMF   & JMARS  & EFM    & MTER   & FacT   &                                                                                             \\ \hline
10                                                                 & 0.0931 & 0.1060 & 0.0564 & 0.1266  & 0.1155 & 0.1071 & 0.1380 & {\bf 0.1499} & 8.62\%*                                                                                      \\ \hline
20                                                                 & 0.1243 & 0.1333 & 0.0825 & 0.1643  & 0.1553 & 0.1354 & 0.1825 & {\bf 0.1991} & 9.10\%*                                                                                      \\ \hline
50                                                                 & 0.1871 & 0.1944 & 0.1345 & 0.2214 & 0.2111 & 0.1903 & 0.2365 & {\bf 0.2488} & 5.20\%*                                                                                      \\ \hline
100                                                                & 0.2509 & 0.2502 & 0.2175 & 0.2668  & 0.2575 & 0.2674 & 0.2783 & {\bf 0.2867} & 3.02\%*                                                                                      \\ \hline
\end{tabular}
\\ * $p$-value < 0.05
\vspace{-3mm}
\end{table*}

\noindent\textbf{$\bullet$ Baselines.} The following popular and state-of-the-art recommendation algorithms are chosen as our baselines for comparison. \\
  \textbf{FMF:} Functional Matrix Factorization \cite{Zhou:2011}. It constructs a decision tree on the user side for tree-based matrix factorization. It was originally designed to solicit interview questions on each tree node for cold-start recommendations. \\
  \textbf{MostPopular (MP):} A non-personalized recommendation solution. Items are ranked by their observed frequency in the training set and the system provides generic recommendations to all users. Though simple, it has shown to be effective in practice \cite{Wang2018}.\\
  \textbf{NMF:} Non-negative Matrix Factorization \cite{Ding:2006}. A widely applied latent factor model for personalized recommendation. \\
  \textbf{BPRMF:} Bayesian Personalized Ranking on Matrix Factorization \cite{Rendle:2009}, which introduces BPR pairwise ranking loss into factorization model learning.\\
  \textbf{JMARS:} A probabilistic model that jointly models aspects, ratings, and sentiments by collaborative filtering and topic modeling \cite{Diao:2014} for explainable recommendation.\\
  \textbf{EFM:} Explicit Factor Models \cite{Zhang2014}. A joint matrix factorization model which constructs user-feature attention and item-feature quality matrices for explainable recommendation. \\
  \textbf{MTER:} Explainable Recommendation via Multi-Task Learning \cite{Wang2018}. A multi-task learning model that integrates user preference modeling for recommendation and opinionated content modeling for explanation via a joint tensor factorization.
  
\noindent\textbf{$\bullet$ Evaluation.} We use Normalized Discounted Cumulative Gain (NDCG) to evaluate the performance of different models. For each dataset, we perform 5-fold cross validation and report the mean value for comparison. Grid search is used to find the optimal hyper parameters in a candidate set for all baseline models, when not explicitly mentioned. 

\subsection{Top-K Recommendation}
We first evaluate \model{}'s recommendation quality. In a good recommender system, items ranked higher in a result list should be more relevant to a user's preference. 
NDCG assigns higher importance to the items ranked on top. In this experiment, we fix the depth of the user tree and item tree in FacT to 6 and the size of latent dimension to 20. The recommendation performance measured by NDCG@{10, 20, 50, 100} of each model is shown in Table~\ref{tab:baselines} for Amazon and Yelp datasets, respectively.

Compared with all the baselines, FacT consistently gives better recommendation in both Amazon and Yelp datasets. 
Among all the baselines, NMF is widely used in practice. 
But it only uses dyadic user-item ratings for model learning. By introducing the pairwise ranking constraint, BPRMF improves greatly comparing to NMF. However, like reported in previous works \cite{Wang2018,Zhang2014}, BPRMF is also limited to the rating information, and cannot utilize the implicit information included in user reviews. By exploiting the review content for recommendation, JMARS and EFM gave explainable recommendation to users with comparable ranking quality with BPRMF, and MTER showed its potential in providing explanations along with decent recommendation quality. However, they are still limited for different reasons.
JMARS maps users, items and features into the same topic space, where the dependency among them is not preserved. Both EMF and MTER model the users and items as individual vectors by matrix or tensor factorization, while FacT clusters users and items into groups (e.g., the intermediate nodes in user and item trees) to take advantage of in-group homogeneity for better latent factor learning. 
The basic intuition here is that the representations of users and items that share the same preferences or feature qualities should be pushed close to each other. And \model{} enforced it by item feature based tree construction. Moreover, the personalized vectors added to the leaf nodes (as discussed in Section \ref{sec:alternative}) distinguish individual users/items, and provide accurate personalized recommendations. 
Besides, we observe that FacT achieves the significant improvement at NDCG@10 (9.70\% against the best baseline on Amazon and 8.62\% on Yelp) and NDCG@20 (8.59\% against the best baseline on Amazon and 9.10\%  on Yelp). This is important for practical recommender system as FacT can provide more accurate recommendations earlier down the ranking list.

Next we will zoom into \model{} to study the effect of several important hyper-parameters in it, including the size of latent dimensions, weight of pairwise ranking loss, tree depth, number of item features, and the inclusion of parent node factors. 

\subsubsection{Latent Dimensions}
\begin{figure}
\includegraphics[width=8.5cm]{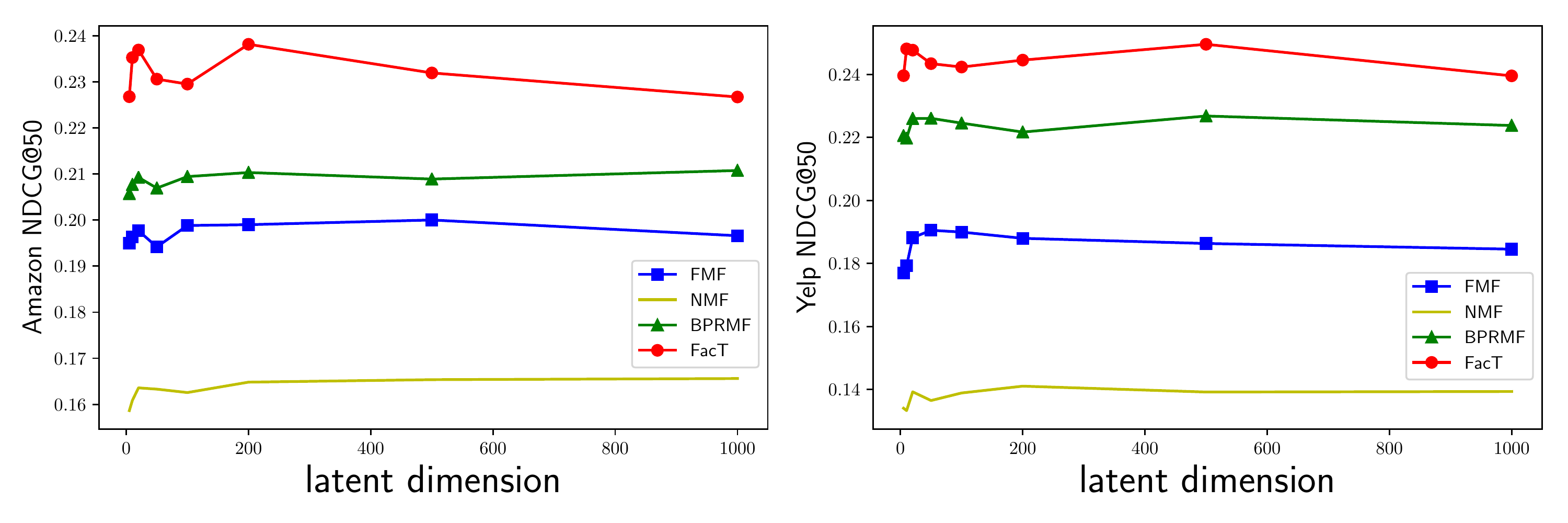}
\vspace{-6mm}
\caption{NDCG@50 v.s., the size of latent dimensions.}
\label{fig-latent_factor}
\vspace{-5mm}
\end{figure}

The dimension of latent factors determines model capacity, and is an important hyper-parameter for factorization based methods.  
In this experiment, we explore the influence of latent dimension and the  stability of \model{} against this hyper-parameter comparing to baseline latent factor models. We varied the dimension of latent factors from 5 to 1000 and compared the results of FacT with FMF, NMF, and BPRMF, which also utilize matrix factorization as their base learning component. The results are summarized in Figure~\ref{fig-latent_factor}.

It is clear from the figure that FacT outperformed the other baselines with NDCG@50 under all different settings. We can also observe that for all models, the performance varied significantly when the dimension was lower. And with larger size of latent dimensions, all models' performance degenerated, as they demand more training data to fit the increasing number of parameters. The situation becomes especially worse for \model{}, as we are also learning latent factors for intermediate nodes. This follows what we expected: it is generally hard for a model with a smaller dimension of latent factors to capture the affinity between users and items; but model with a higher dimension for latent factors is easier to get over-fitted with insufficient training data. Thus, in the following experiment, we choose 20 as the dimension of latent factor in \model{}.

\subsubsection{Contribution of BPR}
Following \cite{Wang2018}, we use the relative normalized weight of BPR to quantify the influence of BPR pairs in FacT and baselines. The relative normalized weight is defined as:

\begin{equation}
    \phi = \frac{\lambda_B \times N_{BPR} \times T_{iter}}{m \times n^2}
\end{equation}
where $N_{BPR}$ is the number of BPR pairs sampled in each iteration and $T_{iter}$ is the number of iterations. Here $m \times n^2$ is the maximum number of all BPR pairs~\cite{Rendle:2009}. In this experiment, we fix $N_{BPR}$, $T_{iter}$ and tune $\lambda_B$ for optimization.

\begin{figure}[t]
\includegraphics[width=8.5cm]{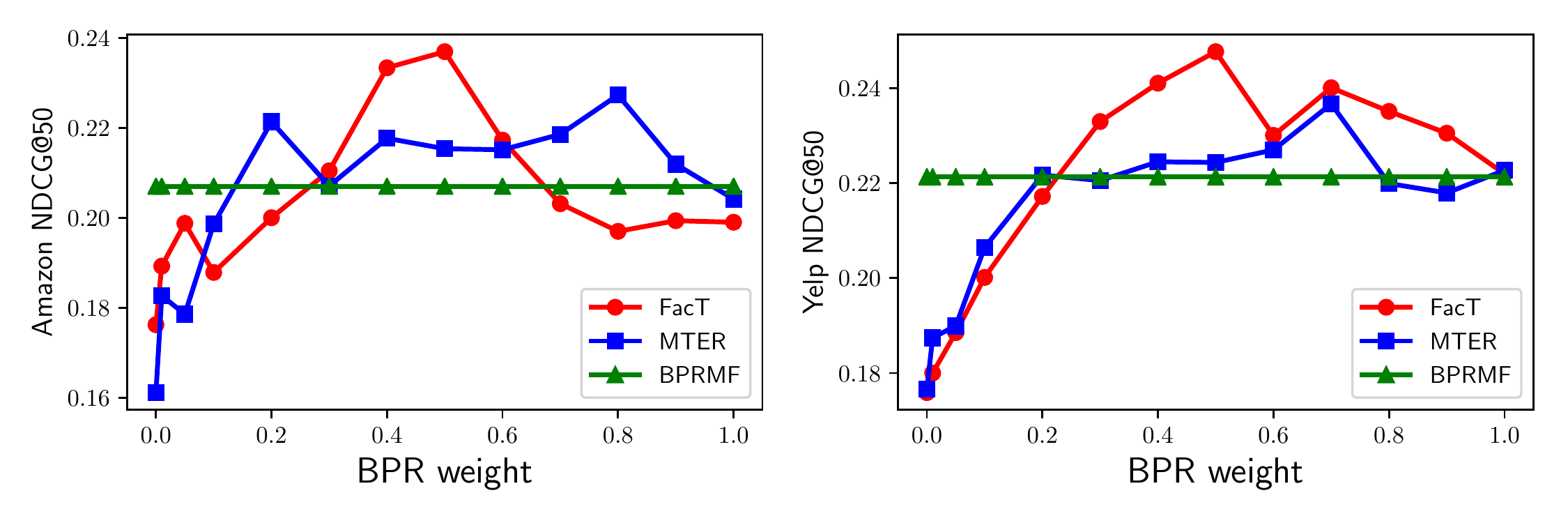}
\vspace{-6mm}
\caption{NDCG@50 v.s. relative BPR weight $\phi$.}
\label{fig-bpr}
\vspace{-5mm}
\end{figure}

The results of NDCG@50 from FacT and two baselines, MTER and BPRMF, are reported in Figure~\ref{fig-bpr}. In this experiment, we varied the BPR weight $\phi$ while keeping all the other hyper-parameters fixed. 
Since BPRMF only optimizes BPR loss, its performance is constant in this experiment. 
It is easy to notice that when $\phi$ is small, the reconstruction error of the training rating matrix dominated both FacT and MTER models, and they performed worse than BPRMF. With an increasing $\phi$, such pairwise loss helped both models identify better latent factors for better ranking. 
However, when $\phi$ further increased, it misled the two models to overfit the BPR loss, and costed a worse ranking quality. When $\phi$ was increased to 1.0, all three models collapsed to almost the same recommendation performance, as the pointwise rating reconstruction loss is totally ignored. 

\subsubsection{Maximum tree depth}
\begin{figure}[t]
  \centering
    \includegraphics[width=8.5cm]{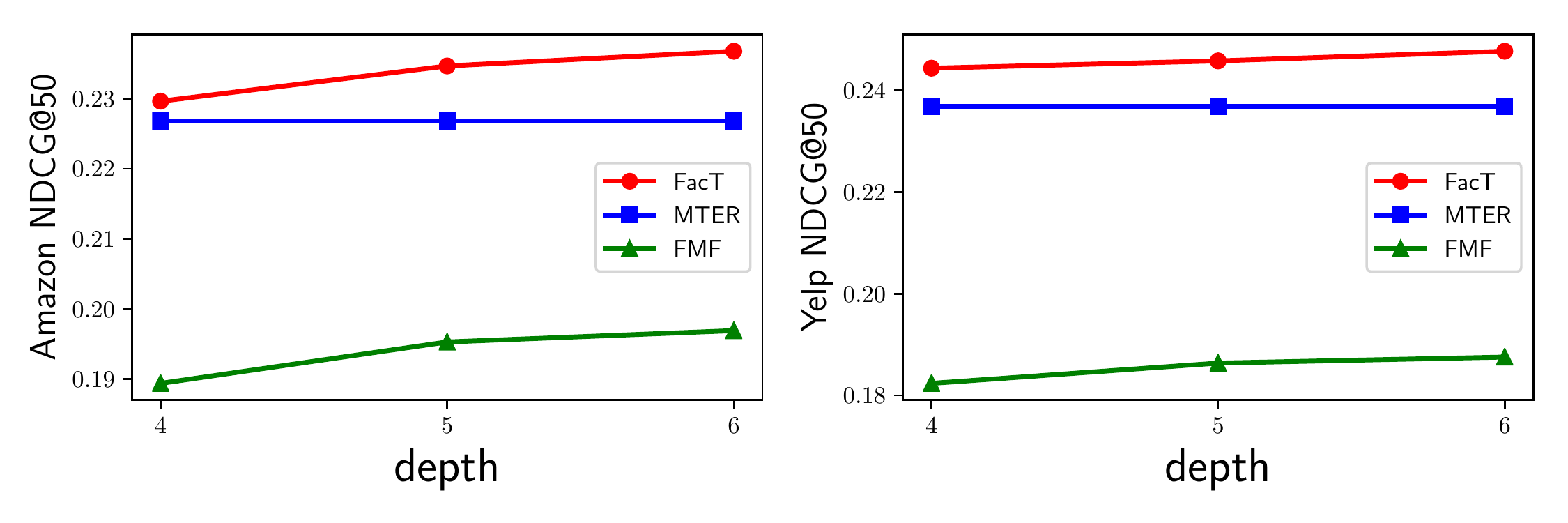}
    \vspace{-6mm}
    \caption{Varying the depth of user tree.}
    \label{fig:user}
\end{figure}

\begin{figure}
    \centering
    \includegraphics[width=9cm]{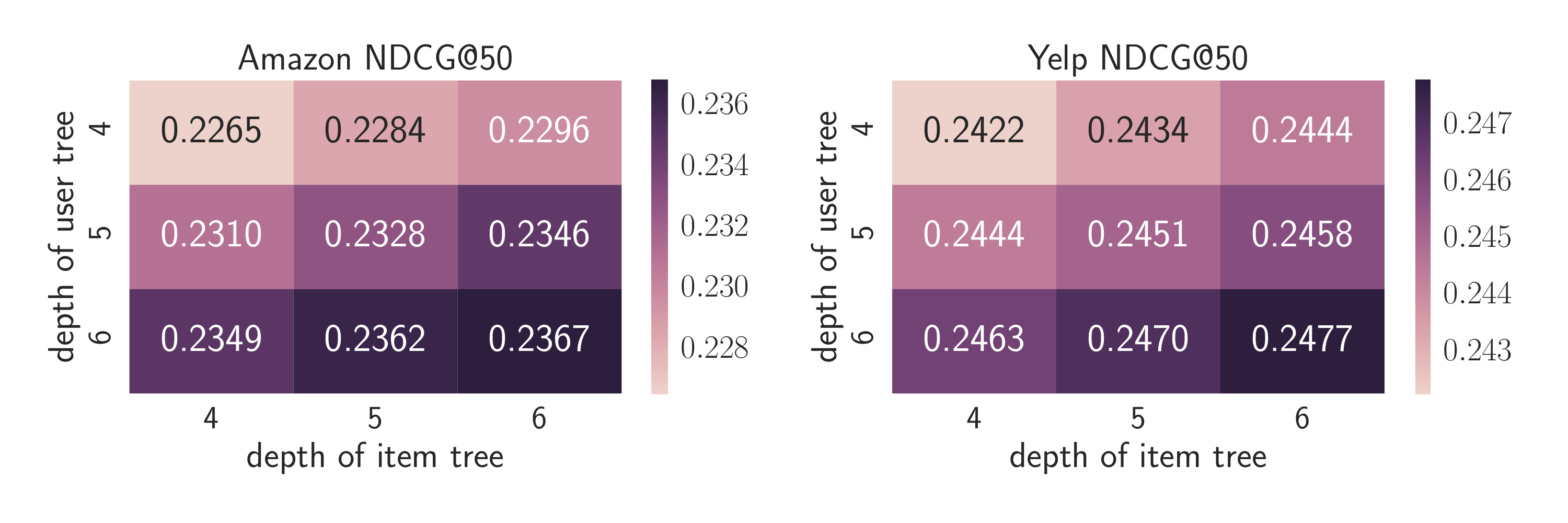}
    \vspace{-6mm}
    \caption{Depth of trees v.s. FacT performance.}
    \vspace{-5mm}
    \label{fig:heatmap}
\end{figure}

In FacT, we cluster the users and items along with the tree construction. The maximum tree depth controls the resolution of clusters, e.g., how many intermediate and leaf nodes will be created. 
We fixed all the other hyper-parameters and only tuned the maximum depth of each tree to verify the effect of it. The results are shown in Figure~\ref{fig:user} and Figure~\ref{fig:heatmap}. In Figure~\ref{fig:user}, 
we compared the performance of FacT with FMF and MTER. FMF introduces user tree construction to cluster users for cold-start recommendation. And MTER is the best baseline we had in Table~\ref{tab:baselines}, but as it does not have a tree structure, its performance remains constant in this experiment. And for FacT, we fixed the depth of item tree to 6 and varied the depth of user tree. We can observe both FMF and FacT got better performance with an increasing tree depth, which increases the granularity of the learnt latent representations for users. 
A more detailed result of varying the depth of both user tree and item tree is shown in Figure~\ref{fig:heatmap}. From this result, we can clearly find that with a larger tree depth, FacT generated consistently better performance. We also notice that the performance change from varying the depth of item tree is much smaller than that from varying the depth of user tree. A possible explanation is that there are only a small portion of items to be recommended to users. The improved resolution on other items has little contribution to \model{}'s ranking quality.

\subsubsection{Number of item features}
\begin{figure}
\includegraphics[width=8.5cm]{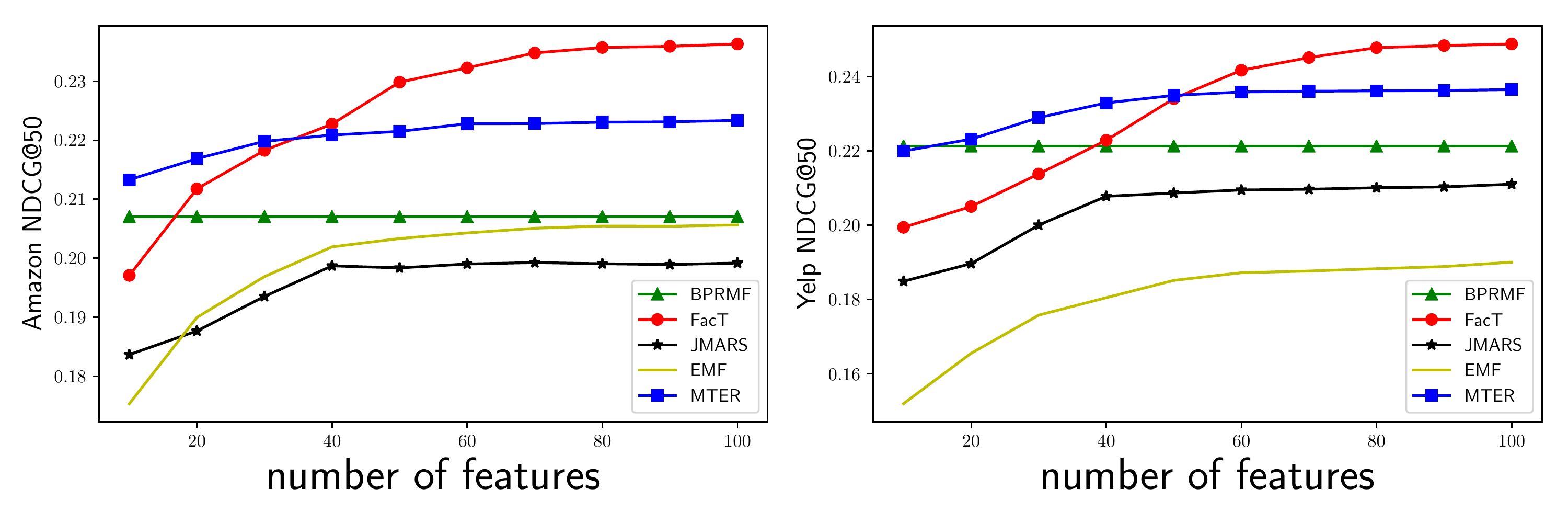}
\caption{NDCG@50 v.s. the number of item features.}
\label{fig-feature}
\end{figure}

As shown in Table~\ref{tab:datasets}, there are 101 item features extracted from Amazon dataset and 104 features from Yelp dataset. Though we have filtered out features with low frequency, limited by the depth of our tree structure, not all of these features will be selected for rule construction. In this analysis, we study the impact of number of features in the dataset on the performance of different models.
We first ordered the features in a descending order of frequency, and then trained the models with an increasing number of features. The results are reported in Figure~\ref{fig-feature}. From Figure~\ref{fig-feature}, it is easy to observe that all the models get significantly improved with an increasing number of features. As the number of features got larger, the performance became stable, as more less frequent features were added. This observation suggests that features with high frequency in reviews contribute more to the feature-based recommendation algorithm learning. Especially in \model{}, when the number of item features is limited, it cannot correctly create tree branches to guide latent factor learning. And more item features give \model{} a higher degree of freedom to recognize the dependency between users and items. 

\subsubsection{Inclusion of factors from parent nodes}
As discussed in Section~\ref{sec:alternative}, during the tree construction, the learnt latent factors from parent nodes are introduced to the latent factor learning of child nodes. Thus, information about homogeneity in grouped users and items could be passed along the tree. 
In this experiment, we quantify the contribution of this design in FacT by disabling it. From Table \ref{tab:personalized}, we can observe that with the personalized term, the model gives significantly better recommendation performance than without it. This directly demonstrates the significance of information sharing among the clustered users and items in \model{}.

\begin{table}[t]
\caption{NDCG@50 for FacT with/without inclusion of factors from parent node. (PF: Parent Factor)}
\vspace{-2mm}
\label{tab:personalized}
\begin{tabular}{|c|c|c|c|c|}
\hline
\multirow{2}{*}{\begin{tabular}[c]{@{}c@{}}Maximum\\ Depth\end{tabular}} & \multicolumn{2}{c|}{Amazon} & \multicolumn{2}{c|}{Yelp} \\ \cline{2-5} 
                                                                         & w/ PF        & w/o PF       & w/ PF       & w/o PF      \\ \hline
4                                                                        & 0.2265       & 0.1811       & 0.2422      & 0.1837      \\ \hline
5                                                                        & 0.2328       & 0.1854       & 0.2451      & 0.1906      \\ \hline
6                                                                        & 0.2367       & 0.1892       & 0.2477      & 0.1985      \\ \hline
\end{tabular}
\vspace{-4mm}
\end{table}

\subsubsection{Dependency on training data}
\begin{figure}[t]
\includegraphics[width=8.5cm]{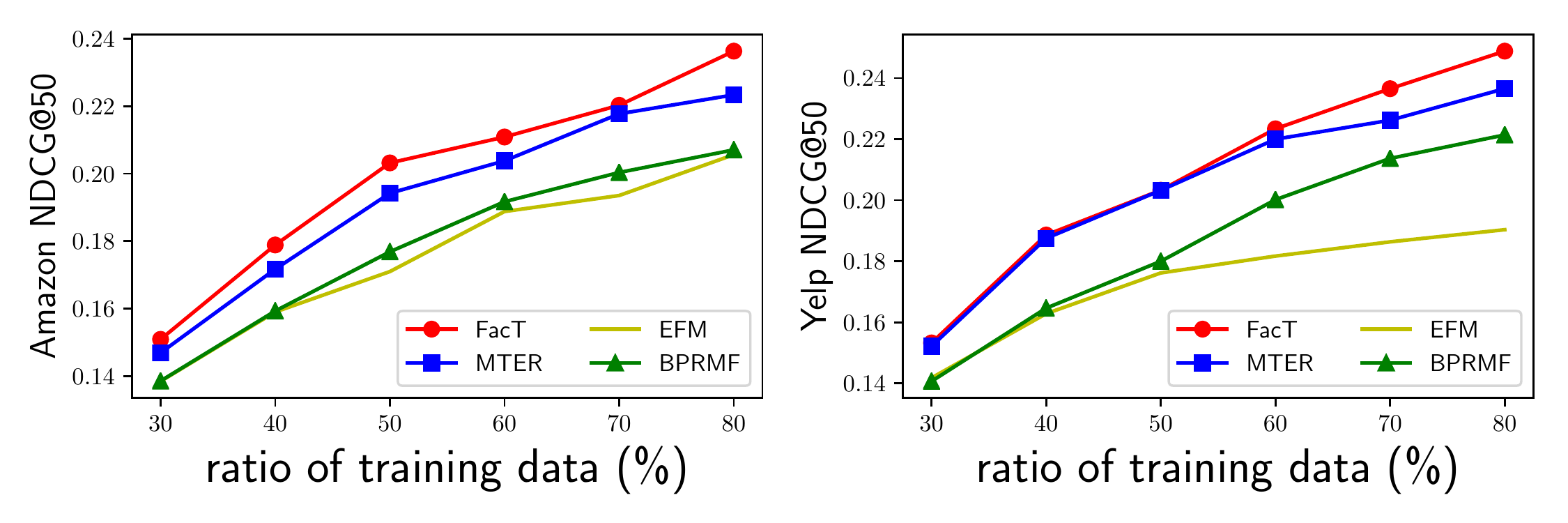}
\vspace{-6mm}
\caption{NDCG@50 v.s. the amount of training data.}
\label{fig-training}
\vspace{-5mm}
\end{figure}

The last thing we investigate is different recommendation algorithms' dependency on the availability of training data. A model requiring less training data is always preferred. We used 30\% to 80\% of training in each fold during 5-fold cross validation in all the models, and reported the results in Figure \ref{fig-training}. As expected all models performed better when more training data became available; by exploiting the shared information across users and items assigned to the same tree node, FacT better utilized the available information and stably outperformed all of the baselines. 

\subsection{Cold-start Recommendation}
Cold-start is an well-known and challenging problem in recommender systems. Without sufficient information about a new user, it is hard for a recommender system to understand the user's interest and provide recommendations with high quality. 
A by-product of \model{} is that the rules learnt in the user tree naturally serve as a set of interview questions to solicit user preferences when a new user comes to the system, i.e., cold-start. For example, based on the user tree in Figure \ref{fig-tree}, the system would get a good understanding of a new user by asking just a few questions following the paths on the tree.
In this experiment, we study how FacT performs on the new users. First, we separated the users into two disjoint subsets, containing 95\% and 5\% users, for training and testing respectively. 
On the training set, we learnt the model and built the user tree and item tree. During testing, for each testing user, we select their first $k$ reviews to construct his/her item feature based user profile (i.e., $F^u_i$ as defined in Eq \eqref{eq_user_profile}). By matching against the user tree, we can easily find the leaf node for each testing user. Then, we use the latent factors reside in the selected leaf node to rank items for this user. We evaluate the performance in the remaining observations from the same user as ground-truth.

\begin{figure}
\includegraphics[width=8.5cm]{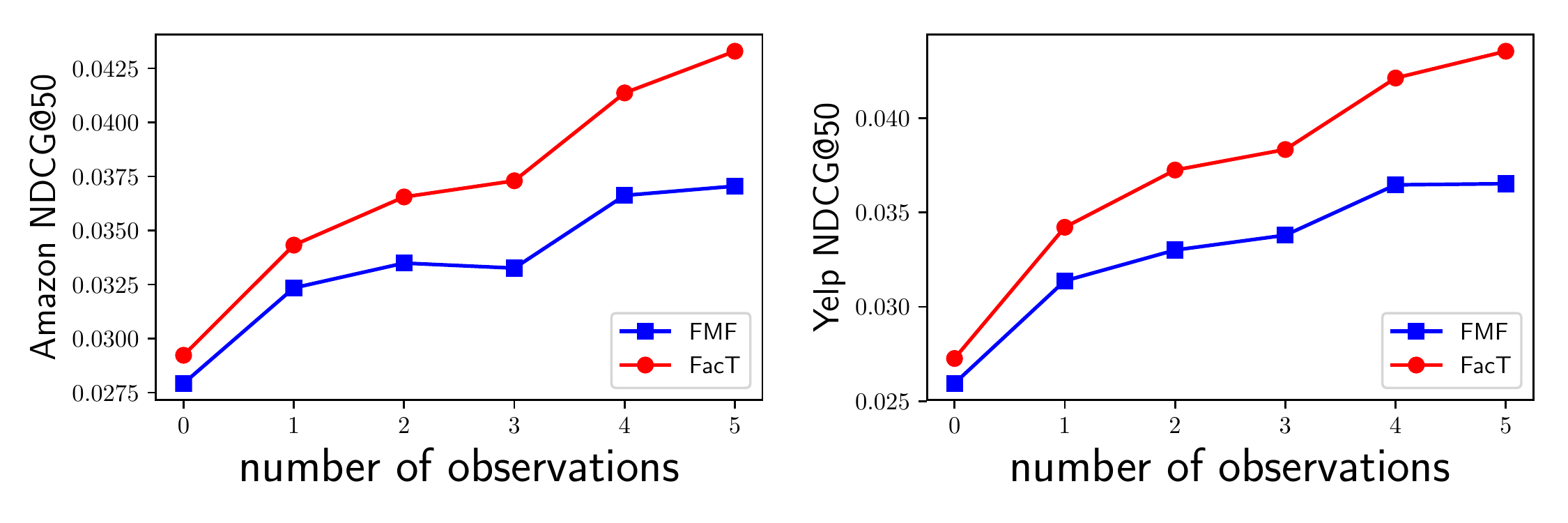}
\vspace{-6mm}
\caption{NDCG@50 v.s., the \# observations in cold-start.}
\vspace{-5mm}
\label{fig-coldstart}
\end{figure}

We compared FacT with FMF model as it is the only baseline that can handle cold-start. We varied the number of observations for each testing user from 0 to 5, and the results are shown in Figure~\ref{fig-coldstart}. First, it is clear to observe that NDCG got improved with an increasing number of observations used to create the user profile for both FacT and FMF. This indicates the effectiveness of user clustering on the user tree in these two models. Second, thanks to the construction of item tree and BPR constraint, FacT got consistently better performance than FMF. In particular, NDCG@50 for FacT increases faster than FMF with more observations. We attribute this to the fact that FacT uses the item features and user opinions collected from the reviews to perform tree construction, while FMF only uses the item ratings to group users. This indicates the effectiveness of review information in modeling users. 

%% file: userstudy.tex
\section{User Study for Explanability}
We performed serious user studies to evaluate user satisfaction of both the recommendations and explanations generated by FacT. We evaluated the performance of FacT on both warm-start users, whose ratings and reviews are known to the system beforehand, and cold-start users who are totally new to the system. The study is based on the review data in Amazon and Yelp datasets used in Section \ref{sec_exp}. We recruited participants on Amazon Mechanical Turk to interact with our system and collected their responses. To reduce the variance and noise of the study, we required the participants to come from an English-speaking country, older than 18 years, and have online shopping experience.    

\subsection{Warm-start Users}
In the warm-start setting, we assume user's purchase history is known to the recommender system. However, we are not able to trace the participants' purchase history on Mechanical Turk. Instead, we performed a simulation-based study, in which we asked the participants to evaluate our system from the perspective of selected users in our datasets. 
Specifically, for each participant, we randomly selected a user from our review dataset and presented this user's reviews for the participant to read. The participants were expected to infer this user's preferences from the review content. Then the participant will be asked several questions to evaluate the recommendation and explanation generated by our algorithm. 
We carefully designed the survey questions to evaluate different aspects of our recommender algorithm as follows:
\begin{itemize}
  \item [Q1:] Generally, are you satisfied with our recommendations?
  \item [Q2:] Do the explanations presented to you really match your preference?
  \item [Q3:] Do you have any idea about how we make recommendations for you?
\end{itemize}
We intended to use {Q1} to evaluate user satisfaction of recommended items, use {Q2} to judge the effectiveness of explanations, and use {Q3} to evaluate the transparency of an explainable recommendation algorithm.
For each question, the participants are required to choose from five rated answers: 1. Strongly negative; 2. Negative; 3. Neutral; 4. Positive; and 5. Strongly positive. We used EFM and MTER as baselines, since they both can provide textual explanations, and conducted A/B tests to ensure the evaluation is unbiased. Three hundred questionnaires were collected in total and the results are reported in Table \ref{tab:warm-start}.
\begin{table}[]
\caption{Result of warm-start user study.}
\label{tab:warm-start}
\vspace{-3mm}
\begin{tabular}{|c|c|c|c|c|c|c|}
\hline
\multirow{2}{*}{Average Score} & \multicolumn{3}{c|}{Amazon}             & \multicolumn{3}{c|}{Yelp} \\ \cline{2-7} 
                               & EFM  & \multicolumn{1}{c|}{MTER} & FacT & EFM     & MTER   & FacT   \\ \hline
Q1                             & 3.64 & 3.96                      & \textbf{4.45*} & 3.45    & 4.06   & \textbf{4.30*}    \\ \hline
Q2                             & 3.48 & 3.88                      & \textbf{4.03} & 3.40     & 3.87   & \textbf{4.13}   \\ \hline
Q3                             & 3.07 & 3.02                      & \textbf{3.88*} & 2.98    & 3.26   & \textbf{3.94*}   \\ \hline
\end{tabular}
\\ * $p$-value < 0.05
\vspace{-5mm}
\end{table}

From  the statistics, \model{} apparently outperformed both baselines in all aspects of this user study, which is further confirmed by the paired t-test. Comparing FacT with EFM and MTER on Q1, the improvement in offline validated recommendation quality directly translated into improved user satisfaction. For Q2, the advantage of FacT shows the effectiveness of our predicate selection in explanation rule construction, which captures user's underlying preferences. Moreover, the results on Q3 verified the user-perceived transparency of our tree guided recommendation and rule-based explanation mechanism.  

\subsection{Cold-start Users}
Unlike warm-start users, cold-start users have no review history. In order to generate recommendation and explanation for these users, we progressively query user responses through an interview process. Specifically, each node of the user tree in \model{} corresponds to an interview question: \textit{"How do you like this [feature]?"}, where \textit{[feature]} was learnt to optimize the explanation rule at this node. When the user answers the interview question designated at the current node, he/she will be directed to one of its three child nodes according to the answer. As a result, each user follows a possibly different path from the root node to a leaf node during the interview process. A user's associated latent factor is adaptively refined at each intermediate node based on the user's responses. We make recommendations and explanations according to the resulting path. For comparison, FMF is set as a baseline, since it is the only algorithm that can address the cold-start problem with the same interview process as FacT. As FMF uses items instead of features to construct the tree, the interview question there is changed to \textit{"How do you like this [item]?"} 

To interview each participant in this user study, we developed a platform to let the participant interact with our system.
\footnote{\href{https://aobo-y.github.io/explanation-recommendation/}{https://aobo-y.github.io/explanation-recommendation/}}
To increase the sensitivity of comparison between two recommendation algorithms, we conduct interleaved test \cite{Radlinski:2008:CDR:1458082.1458092} in this cold-start study. The participant was asked to interact with two models one after the other in a random order, to compare which one is better according to our designed questions. 
The recommendation is interactive, based on the participants' responses to the interview questions (i.e., traversing in the user tree).  
There are three questions for them to answer to compare the recommendations and explanations generated by these two algorithms:
\begin{itemize}
  \item [Q1:] Generally, between system A and B, whose recommendations are you more satisfied with?
  \item [Q2:] Between system A and B, whose explanations do  you think can better help you understand the recommendations?
  \item [Q3:] Between system A and B, whose explanations can better help you make a more informed decision?
\end{itemize}
We collected more than 100 valid responses on each dataset and reported the results in Table \ref{tab:cold-start}.

We can find that \model{} is preferred than FMF in all questions on both datasets. It suggests that: First, feature-based rule construction is more effective than item-based rule construction, which leads to improved  ranking quality in \model{}. Second, the feature-based explanations are preferred than the item-based ones, as the former  characterizes user preferences at a finer granularity. Last, feature-based explanation rules also provide improved transparency than item-based explanations, which verifies the explainability of our solution. All the evidences from this interleaved user study demonstrate the power of FacT to address the cold-start  problem.

\begin{table}[]
\caption{Results of cold-start interleaved test.}
\vspace{-3mm}
\label{tab:cold-start}
\begin{tabular}{|c|c|c|c|c|}
\hline
\multirow{2}{*}{number of votes} & \multicolumn{2}{c|}{Amazon}     & \multicolumn{2}{c|}{Yelp} \\ \cline{2-5} 
                                 & FMF & \multicolumn{1}{l|}{FacT} & FMF         & FacT        \\ \hline
Q1                               & 44  & \textbf{63*}                        & 40          & \textbf{64*}          \\ \hline
Q2                               & 43  & \textbf{64*}                        & 34          & \textbf{70*}          \\ \hline
Q3                               & 45  & \textbf{62}                        & 33          & \textbf{71*}          \\ \hline
\end{tabular}
\\ * $p$-value < 0.05
\vspace{-5mm}
\label{tab:cold-start}
\end{table}